\documentclass[11pt,a4paper]{article}
\usepackage{jheppub}
\usepackage[utf8]{inputenc}
\usepackage{physics}
\usepackage{tikz-feynman}
\usepackage{caption}
\usepackage{xcolor}
\usepackage{comment}
\usepackage{multirow}
\usepackage{graphics}
\usepackage{float}
\usepackage{cancel}
\usepackage{soul}
\usepackage{cancel}
\usepackage{array}

\tikzfeynmanset{compat=1.1.0}
\newcolumntype{L}[1]{>{\raggedright\let\newline\\\arraybackslash\hspace{0pt}}m{#1}}
\newcolumntype{C}[1]{>{\centering\let\newline\\\arraybackslash\hspace{0pt}}m{#1}}
\newcolumntype{R}[1]{>{\raggedleft\let\newline\\\arraybackslash\hspace{0pt}}m{#1}}

\def\nn{\nonumber}
\def\l{\left}
\def\r{\right}

\title{\boldmath{Bell violation in $2\rightarrow 2$ scattering in photon, gluon and graviton EFTs}}

\author[a,1]{Diptimoy Ghosh}
\author[a,2]{Rajat Sharma}

\affiliation[a]{Department of Physics\\Indian Institute of Science Education and Research Pune, India}

\emailAdd{diptimoy.ghosh@iiserpune.ac.in}
\emailAdd{rajat.sharma@students.iiserpune.ac.in}

\abstract{In this paper, we explore Bell inequality violation
for $2\rightarrow2$ scattering in Effective Field Theories (EFTs)
of photons, gluons, and gravitons. Using the CGLMP Bell parameter ($I_2$), we show that, starting from an appropriate initial non-product state, the Bell inequality can always be violated in the final state (i.e.,$I_2 >2$) at least for some scattering angle. For an initial product state, we demonstrate that abelian gauge theories behave qualitatively differently than non-abelian gauge theories (or Gravity) from the point of view of Bell violation in the final state: in the non-abelian case, Bell violation ($I_2>2$) is never possible within the validity of EFTs for weakly coupled UV completions. Interestingly, we also find that, for a maximally entangled initial state, scattering can reduce the degree of entanglement only for CP-violating theories. Thus Bell violation in $2\rightarrow2$ scattering can, in principle, be used to classify CP conserving vs violating theories.}

\begin{document}

\maketitle

\section{Introduction}
Entanglement is a unique relationship between two or more particles, where their states are correlated in such a way that a measurement performed on one particle instantly influences the other particle, regardless of the spatial separation between them. This correlation poses a challenge to the principle of local realism, which asserts that the properties of a state are determined by its local environment. In 1964 \cite{Bell}, John Bell derived a set of inequalities - commonly known as Bell inequalities - for the correlated expectation values that must be satisfied by a local deterministic theory. However, quantum mechanics predicts that the Bell inequalities can be violated for certain correlated expectation values, which has also been experimentally verified \cite{PhysRevLett.115.250401, PhysRevLett.28.938, PhysRevLett.49.1804, PhysRevLett.115.250402,1206.2031,1508.05949}. These experiments have played a critical role in establishing quantum mechanics as a fundamental theory of nature. 

Entanglement can have significant implications for our understanding of spacetime and information in quantum field theory (QFT) \cite{2201.13310,1803.04993}. However, very little is known and explored about the Bell inequalities in the context of QFT \cite{0912.1941}. Recently, the interest has been revived after it has been shown that the Bell inequalities can be violated experimentally by the entangled top-quark pairs produced at the LHC \cite{2003.02280,2102.11883,2110.10112,2205.00542,2203.05582}. More work along this line has shown that it is possible to experimentally measure Bell violation for hyperons \cite{2107.13007} and gauge bosons from Higgs boson decay \cite{2106.01377,2204.11063} as well.

These observations have motivated the study of entanglement in $2\rightarrow 2$ scattering in high energy physics \cite{1703.02989,2108.00646,1812.03138}, particularly in the context of Effective Field Theory (EFT)\cite{2208.11723,2203.05619,sinha_10.48550,2210.09330}. The existence of higher dimensional operators in an EFT can modify the degree of entanglement in the final scattered states, which might act as a possible probe of new physics. If we experimentally observe Bell violation in $2\rightarrow2$ scattering for a particular initial state, then we can directly constrain the corresponding EFT by quantifying the degree of entanglement in the final scattered state. However, a priori, it is not very clear which initial state can be used to probe the quantum nature of the theory and demand Bell violation.

In this work, we consider the CGLMP Bell parameter ($I_2$) \cite{040404} as the measure of entanglement in the states. For local hidden variable theories, $|I_2|\leq2$, however, this inequality can be violated by quantum theories as shown in section \ref{sec2}. Therefore, the CGLMP Bell parameter can be used to distinguish between the local hidden variable theories and the quantum theories. We consider the initial states for $2\rightarrow2$ scattering such that the CGLMP parameter corresponding to them ($I_{2i}$) satisfies $|I_{2i}|\leq2$. In other words, the CGLMP parameter for the initial state can, in principle, be explained by a local hidden variable theory, and it does not call for a quantum mechanical origin. We use this condition with the motivation that we want to probe the quantum nature of a theory only through the scattering process, i.e., whether unitary evolution can increase the degree of entanglement beyond $I_2=2$. We then calculate the allowed EFT parameter space for which we can observe Bell violation at some energy (within the validity of the EFT regime) and scattering angle.
\vspace{2mm}

The rest of the paper is organized as follows: In section \ref{sec2}, we give a brief overview of the CGLMP Bell parameter and its validity. In section \ref{sec3}, we explore the possibility of Bell violation in photons, gluons, and graviton EFTs. Section \ref{CP_vs_CPv} discusses CP violation from the point of view of Bell violation. The summary of our results is presented in section \ref{summary}.


\section{CGLMP Bell parameter}\label{sec2}
We will first briefly overview the CGLMP inequality and the corresponding Bell parameter for qubits following \cite{040404}. Let us suppose there are two parties, Alice ($A$) and Bob ($B$), with a qubit state each. Alice can carry out two possible measurements, $A_1$ or $A_2$, and Bob can carry out two possible measurements, $B_1$ or $B_2$. Each measurement can have only two possible outcomes: $A_1, A_2, B_1, B_2 = 0,1$. For a local hidden variable theory, the qubit system can be described by $16$ probabilities $c_{jklm}$ where $(j,k)$ are Alice's local variables and $(l,m)$ are Bob's local variables. The pair $(j,k)$ represents that measurement $A_1$ has outcome $j$ and measurement $A_2$ has outcome $k$; and $(l,m)$ represents that the measurement $B_1$ has outcome $l$ and measurement $B_2$ has outcome $m$. Since $c_{jklm}$ are probabilities, they are positive $\l(c_{jklm}\geq0\r)$ and sum to one $\l(\sum_{jklm}c_{jklm}=1\r)$. The joint probabilities $P(A_1=j,B_2=m)$ then take the following form $P(A_1=j,B_2=m)=\sum_{kl}c_{jklm}$ and similarly for $P(A_1=j,B_1=l)$, $P(A_2=k,B_1=l)$ and $P(A_2=k,B_2=m)$. Using the joint probabilities, we define the probability $P(A_a=B_b+k)$ that the measurements $A_a$ and $B_b$ have outcomes that differ by $k$ modulo $2$ :
\begin{equation}\label{probab_k}
    P(A_a=B_b+k)\equiv\sum_{j=0}^{1}P(A_a=j, B_b=j+k \text{ mod } 2)
\end{equation}
The CGLMP inequality\footnote{For the qubit case, the CGLMP inequality is equivalent to the CHSH inequality \cite{CHSH}.} is a combination of the above probabilities defined as:
\begin{align}
    I_2=&+[P(A_1=B_1)+P(B_1=A_2+1)+P(A_2=B_2)+P(B_2=A_1)]\\\nn
    &-[P(A_1=B_1-1)+P(B_1=A_2)+P(A_2=B_2-1)+P(B_2=A_1-1)]
\end{align}
For a local hidden variable theory, we can have any three probabilities with a $+$ sign in the above expression satisfied along with one with a $-$ sign (or vice versa). Therefore, for such theories $-2\leq I_2\leq2$. However, for a quantum mechanical theory, $I_2$ can be greater than 2, as shown below.\\

\noindent Consider the following normalized quantum state of entangled qubits (in $\ket{0}\otimes\ket{0}$ and $\ket{1}\otimes\ket{1}$ basis),
\begin{equation}
    \ket{\Psi}=\frac{1}{\sqrt{\sum_{m=0}^1|\mu_{m}|^2}}\sum_{m=0}^1\mu_{m}\ket{m}_A\otimes\ket{m}_B
\end{equation}
The measurements by Alice and Bob are carried out in three steps \cite{040404,0101084}. First, a variable phase, $e^{i\phi_a(m)}$ for Alice and $e^{i\varphi_b(m)}$ for Bob, which depends on the measurement being carried out is given to each state $\ket{m}$ using phase shifters which are at the disposal of the observer. Thus the state becomes 
\begin{equation}
    \ket{\Psi}=\frac{1}{\sqrt{\sum_{m=0}^1|\mu_{m}|^2}}\sum_{m=0}^1\mu_{m}e^{i\phi_a(m)}e^{i\varphi_b(m)}\ket{m}_A\otimes\ket{m}_B
\end{equation}
where $\phi_1(m)=\pi\alpha_1 m$, $\phi_2(m)=\pi\alpha_2 m$, $\varphi_1(m)=\pi\beta_1 m$ and $\varphi_2(m)=\pi\beta_2 m$ with $\alpha_1=0$, $\alpha_2=1/2$, $\beta_1=1/4$ and $\beta_2=-1/4$. These are the optimal measurement settings for which one gets the maximum value of $I_2$ for an entangled quantum state \cite{040404}.

Then each party carries out a discrete Fourier transform to get the state to the following form,
\begin{equation}
    \ket{\Psi}=\frac{1}{2\sqrt{\sum_{m=0}^1|\mu_{m}|^2}}\sum_{m,k,l=0}^1\mu_{m}\text{exp}\bigg[i\Big(\phi_a(m)+\varphi_b(m)+\pi m(k-l)\Big)\bigg]\ket{k}_A\otimes\ket{l}_B
\end{equation}
The final step is for Alice to measure the projection of the state along the $k$ basis and for Bob to measure along the $l$ basis. Thus the joint probabilities are:
\begin{equation}\label{joint_probab}
    P(A_a=k,B_b=l)=\frac{1}{4\sum_{m=0}^1|\mu_{m}|^2}\left|\sum_{m=0}^1\mu_{m}\text{exp}\bigg[i\pi m\Big(\alpha_a+k+\beta_b-l\Big)\bigg]\right|^2
\end{equation}
We can also get the above joint probabilities in one step if we consider the operators $A_a$ and $B_b$ to have the following non-degenerate eigenvectors, respectively \cite{sinha_10.48550},
\begin{align}
|k\rangle_{A, a} & =\frac{1}{\sqrt{2}} \sum_{j=0}^{1} X_{j, k}^{(a)}|j\rangle_A, \\
|l\rangle_{B, b} & =\frac{1}{\sqrt{2}} \sum_{j=0}^{1} Y_{j, l}^{(b)}|j\rangle_B,
\end{align}
where $X_{j, k}^{(a)}=\exp \left(i\pi j\left(k+\alpha_a\right)\right), Y_{j, l}^{(b)}=\exp \left(i \pi j\left(-l+\beta_b\right)\right)$ with $\alpha_a$ and $\beta_b$ defined as before. Then we get the following joint probabilities, 
\begin{equation}
P\left(A_a=k, B_b=l\right)=\bra{\Psi}\left(\ket{k}_{A, a} \otimes\ket{l}_{B, b} \quad {}_{A, a}\bra{l}\otimes_{B, b}\bra{k}\right) \ket{\Psi}
\end{equation}
which when evaluated is same as $\text{eq}^{n}$(\ref{joint_probab}).\\
Using $\text{eq}^\text{n}$(\ref{joint_probab}) and $\text{eq}^\text{n}$(\ref{probab_k}), we can calculate $I_2$ for generic $\mu_{m}$, which is given by
\begin{equation}\label{CGLMP}
    I_2=\frac{2\sqrt{2}\l(\mu_0\mu_1^*+\mu_1\mu_0^*\r)}{\sqrt{\sum_{m=0}^1|\mu_{m}|^2}}
\end{equation}
When $I_2$ is extremized w.r.t $\mu_{m}$, we get $-2\sqrt{2}\leq I_2\leq 2\sqrt{2}$. The $I_2 = 2\sqrt{2}$ corresponds to the maximally entangled state, $\ket{\Psi}=\l(\ket{0}\otimes\ket{0}+\ket{1}\otimes\ket{1}\r)/\sqrt{2}$. Thus for a quantum mechanical theory, $|I_2|$ can be greater than 2, whereas $|I_2|$ is always less than or equal to 2 for a local hidden variable theory. In the rest of the paper, we will denote $\ket{m}\otimes\ket{n}$ as $\ket{m,n}$ for convenience.\\

Note that if we consider $\ket{\Psi}$ to be a generic superposition in $\ket{0,0}$, $\ket{0,1}$, $\ket{1,0}$ and $\ket{1,1}$ basis,
\begin{equation}\label{final_psi}
    \ket{\Psi}=\sum_{m,n=0}^1\mu_{m,n}\ket{m,n}
\end{equation}
then we get the following expression for $I_2$,
\begin{equation}
    I_2=\frac{2\sqrt{2}\l(\mu_{0,0}\mu_{1,1}^*+\mu_{1,1}\mu_{0,0}^*\r)}{\sqrt{\sum_{m,n=0}^1|\mu_{m,n}|^2}}
\end{equation}
However, now we get $I_2=0$ for certain maximally entangled states, $\ket{\Psi}=(\ket{0,1}+\ket{1,0})/\sqrt{2}$ and $\ket{\Psi}=(\ket{0,0}+i\ket{1,1})/\sqrt{2}$. Therefore, $I_2$ is a good order parameter for entanglement only if one considers $\ket{0,0}$ and $\ket{1,1}$ as the basis and real coefficients $\mu_0$ and $\mu_1$. For this purpose, we go to $\ket{0,0}$ and $\ket{1,1}$ basis before calculating $I_2$ 
 and only consider CP-conserving theories as they have real amplitudes.
%
\section{Bell inequality in $2\rightarrow 2$ scattering}\label{sec3}

Since Bell inequalities can be used to distinguish between local hidden variable theories and quantum theories, we try to probe the quantum nature of different EFTs using $2\rightarrow2$ scattering and Bell inequalities. For this purpose, we relate the CGLMP Bell parameter $I_2$ to the $2\rightarrow 2$ scattering helicity amplitudes, which can be easily calculated. 

Consider $\ket{p_1,h_1,m_1;p_2,h_2,m_2}$ and $\ket{p_3,h_3,m_3;p_4,h_4,m_4}$ to be the basis for initial and final states, respectively, for the scattering process. Here, $h_i$ represents the helicity of the particles, which can take values $h=+1,-1$ for photons and gluons, and $h=+2,-2$ for gravitons; and $m_i$ represents the other quantum numbers, like color for gluons. We sum over the other quantum numbers for final states, $m_3$ and $m_4$, so that the final states are entangled only in Hilbert space spanned by helicity states. It would also be interesting to study the entanglement in other quantum numbers, however, one has to consider the appropriate CGLMP Bell parameter, $I_d$ \cite{040404}.

Now, the final states can be represented as $\ket{p_3,h_3;p_4,h_4}$. We denote the helicity scattering amplitudes as $\mathcal{M}^{h_3,h_4}_{h_1,h_2}(m_1,m_2,s,t,u)$ where we take initial states to be incoming and final states to be outgoing and s,t,u are the usual Mandelstam variables. In the rest of the paper, we use $1$ in place of $h=+1,+2$ and 0 in place of $h=-1,-2$, for convenience.

Consider the initial state to be a generic superposition in the $\ket{0,0}$ and $\ket{1,1}$ helicity basis, $\ket{\psi}_i=\sum_{i=0,1}\lambda_{i}\ket{i,m_1;i,m_2}$. We have suppressed $p_i$ here and in the rest of the paper to make the notation less clumsy. Any state can be written in this basis by the Schmidt decomposition theorem, as also shown in the appendix. We consider the initial states such that the CGLMP parameter corresponding to them ($I_{2i}$) satisfies $|I_{2i}|\leq2$ as we want to probe quantum nature of the theory only through the scattering process.
We then identify $\mu_{m,n} = \sum_{i=0,1}\lambda_{i}\mathcal{M}_{i,i}^{m,n}$ in $\text{eq}^\text{n}$(\ref{final_psi}) i.e. we take $\ket{\Psi}$ to be the final state of our scattering process.
%
We convert the final state to $\ket{0,0}$ and $\ket{1,1}$ basis by the Schmidt decomposition method and calculate $I_{2f}$ for different theories and initial states using $\text{eq}^\text{n}$(\ref{CGLMP}). \\

\vspace{-3mm}
\noindent From now on, we will use the parameterization $\lambda_0\rightarrow\cos\theta$
 and $\lambda_1\rightarrow\sin\theta$ for convenience.
 \vspace{2mm}

\noindent We consider the scattering of identical particles and assume parity symmetry. Then for photons, gluons, and gravitons scattering (cases that we will be considering), there are total 16 helicity scattering amplitudes for each set of additional quantum numbers $m_1$ and $m_2$ (they represent colors for gluons and do not exist for photons and gravitons). Note that we sum over $m_3$ and $m_4$ while calculating amplitudes, as already mentioned. However, due to parity symmetry, there are only five distinct scattering amplitudes in the COM frame \cite{2211.05795}, denoted as
\begin{align}
    \Phi_1(s, t, u) \equiv \mathcal{M}_{1,1}^{1,1}(s, t, u),\quad & \Phi_2(s, t, u) \equiv \mathcal{M}_{1,1}^{0,0}(s, t, u), \quad \Phi_3(s, t, u) \equiv \mathcal{M}_{1,0}^{1,0}(s, t, u) \\
    \Phi_4(s, t, u) \equiv \mathcal{M}_{1,0}^{0,1}(s, t, u),\quad & \Phi_5(s, t, u) \equiv \mathcal{M}_{1,1}^{1,0}(s, t, u)
\end{align}
where we have suppressed the $m_i$ dependence of helicity amplitudes. If we don't have quantum numbers other than $h_i$'s i.e. $m_i$'s  do not exist, as is the case for photons and gravitons, then due to crossing symmetry, $\Phi_3$ and $\Phi_4$ can be related to $\Phi_1$ as,
$$\Phi_3(s, t, u)=\Phi_1(u, t, s), \quad \Phi_4(s, t, u)=\Phi_1(t, s, u),$$
This leads to only three independent helicity amplitudes, $\Phi_1$, $\Phi_2$ and $\Phi_5$. All the helicity scattering amplitudes for CP conserving theories can be denoted as,
\begin{equation}\label{CPamp}
\left(\begin{array}{cccc}
\mathcal{M}_{1,0}^{1,0} & \mathcal{M}_{1,0}^{1,1} & \mathcal{M}_{0,0}^{1,0} & \mathcal{M}_{0,0}^{1,1} \\
\mathcal{M}_{1,0}^{0,0} & M_{1,0}^{0,1} & M_{0,0}^{0,0} & M_{0,0}^{0,1} \\
M_{1,1}^{1,0} & \mathcal{M}_{1,1}^{1,1} & \mathcal{M}_{0,1}^{1,0} & \mathcal{M}_{0,1}^{1,1} \\
\mathcal{M}_{1,1}^{0,0} & \mathcal{M}_{1,1}^{0,1} & \mathcal{M}_{0,1}^{0,0} & \mathcal{M}_{0,1}^{0,1}
\end{array}\right)=\left(\begin{array}{cccc}
\Phi_3 & \Phi_5 & \Phi_5 & \Phi_2 \\
\Phi_5 & \Phi_4 & \Phi_1 & \Phi_5 \\
\Phi_5 & \Phi_1 & \Phi_4 & \Phi_5 \\
\Phi_2 & \Phi_5 & \Phi_5 & \Phi_3
\end{array}\right)
\end{equation}
We work in mostly \textit{minus} signature, $\eta_{\mu\nu}=(+,-,-,-)$ throughout our work.

\subsection{Euler-Heisenberg}
Let's consider the following EFT Lagrangian for photons up to dim 8,
\begin{equation}
    L=-\frac{1}{4}F_{\mu\nu}F^{\mu\nu}+\frac{c_1}{\Lambda^4}(F^{\mu\nu}F_{\mu\nu})^2+\frac{c_2}{\Lambda^4}(F^{\mu\nu}\widetilde{F}_{\mu\nu})^2
\end{equation}
For the case of photons, we don't have to worry about the other quantum numbers, $m_1$ and $m_2$, and the low energy amplitudes $\Phi_1, \Phi_2$, and $\Phi_5$ can be written as 
\begin{align}\label{photon_amp}
& \Phi_1(s, t, u)=g_2 s^2 \\\nn
& \Phi_2(s, t, u)=f_2\left(s^2+t^2+u^2\right)\\\nn
& \Phi_5(s, t, u)=0
\end{align}
up to $\mathcal{O}\left(1/\Lambda^4\right)$ \cite{2211.05795}, where $g_2=8(c_1+c_2)/\Lambda^4$ and $f_2=8(c_1-c_2)/\Lambda^4$.
%
\vspace{2mm}

\noindent For a generic initial state, $\ket{\psi}_i=\cos\theta\ket{0,0}+\sin\theta\ket{1,1}$, we get the following final state after $2\rightarrow2$ scattering,
\begin{align}
      \ket{\Psi}=&\cos\theta(\mathcal{M}_{0,0}^{0,0}\ket{0,0}+\mathcal{M}_{0,0}^{1,1}\ket{1,1}+\mathcal{M}_{0,0}^{0,1}\ket{0,1}+\mathcal{M}_{0,0}^{1,0}\ket{1,0})\\\nn
      &\sin\theta(\mathcal{M}_{1,1}^{0,0}\ket{0,0}+\mathcal{M}_{1,1}^{1,1}\ket{1,1}+\mathcal{M}_{1,1}^{0,1}\ket{0,1}+\mathcal{M}_{1,1}^{1,0}\ket{1,0})
\end{align}
Using $\text{eq}^{\text{n}}$(\ref{CPamp}) and $\text{eq}^{\text{n}}$(\ref{photon_amp}), the final state can be written as
\begin{align}
\ket{\Psi}=(\sin\theta\;\Phi_1+\cos\theta\;\Phi_2 )\ket{1,1}+(\sin\theta\;\Phi_2+\sin\theta\;\Phi_1 )\ket{0,0}
\end{align}
Since the above final state is already in $\ket{0,0}$ and $\ket{1,1}$ basis, we can directly calculate the corresponding CGLMP Bell parameter $I_{2f}$ using $\text{eq}^{\text{n}}$(\ref{CGLMP}),
\begin{align}\label{I2f_photons}
    I_{2f}&=\frac{2\sqrt{2}\l(2\Phi_1\Phi_2+(\Phi_1^2+\Phi_2^2)\sin2\theta\r)}{|\Phi_1|^2+|\Phi_2|^2+2\Phi_1\Phi_2\sin2\theta}\\\nn
    &=\frac{2\sqrt{2}\l(\sin2\theta+4f^2(1+\chi+\chi^2)^2\sin2\theta+4f(1+\chi+\chi^2)\r)}{1+4f^2(1+\chi+\chi^2)^2+4f(1+\chi+\chi^2)\sin2\theta}
\end{align}
where $f=f_2/g_2$ and $-1\leq\chi=t/s\leq0$ for physical $t$ and $s$. Also, $\cos\phi_s=1+2\chi$ where $\phi_s$ is the scattering angle.

We extremize $I_{2f}$ w.r.t $\theta$ and $\chi$ with the constraint $|I_{2i}|\leq2$. We find that one can observe Bell violation, i.e., $|I_{2f}|>2$, for some scattering angle (or equivalently $\chi$) for all values of $f_2$ and $g_2$ except for $f_2=0$ or $g_2=0$. In other words, for any non-zero values of $f_2$ and $g_2$, one can observe Bell violation at some scattering angle if $\ket{\psi}_i$ is chosen appropriately. Since we are interested in exploring the possibility of Bell violation due to the quantum evolution of the initial state dictated by the theory, we take the initial state whose CGLMP Bell parameter ($I_{2i}$) is less than 2, i.e., it can, in principle, also be described by a local hidden variable theory.

For $f_2=0$ or $g_2=0$, the maximum value for $I_{2f}$ is equal to $2$, with the constraint $|I_{2i}|\leq2$, which lies on the boundary of Bell inequality. Therefore, for $|c_1|=|c_2|$, there is no Bell violation for any value of $\chi$ and $|I_{2i}|\leq2$.
%
\vspace{2mm}

\noindent \textbf{Product initial state:} If we take the product state $\ket{1,1}$) to be the initial state in $2\rightarrow2$ scattering (i.e. $\theta=\pi/2$ in $\ket{\psi}_i=\cos\theta\ket{0,0}+\sin\theta\ket{1,1}$), instead of the general state with the constraint $|I_{2i}|\leq2$, then we get the following $I_{2f}$
\begin{equation}
    I_{2f}=I_2^{1,1}=\frac{8 \sqrt{2} f_2 g_2\left(\chi^2+\chi+1\right)}{4\left(\chi^2+\chi+1\right)^2 f_2^2+g_2^2}
\end{equation}
For this particular initial state, we observe Bell violation for some scattering angle, given
%
\begin{equation}\label{QED_con}
    \frac{\sqrt{2}-1}{2} \leq \left|\frac{f_2}{g_2}\right| \leq \frac{2(\sqrt{2}+1)}{3} \quad \approx \quad  0.2071\leq \left|\frac{f_2}{g_2}\right| \leq 1.6095
\end{equation}
Interestingly, the QED 1-loop answer for $\left|\frac{f_2}{g_2}\right|\approx 0.2727$ \cite{1810.06994,Heisenberg,1202.1557} lies inside the above range. Note that a similar exercise was done in \cite{sinha_10.48550} by demanding Bell violation for all scattering angles, which leads to a slightly different range for $|f_2/g_2|$.

For $|f_2/g_2|$ outside the above range (\ref{QED_con}), we don't observe Bell violation for any scattering angle because here we have fixed $\theta=\pi/2$ in the initial state. If we allow $\theta$ to vary, then we can observe Bell violation for all non-zero $f_2$ and $g_2$, as shown above ($\text{eq}^\text{n}$(\ref{I2f_photons}) and the following paragraph).

If one performs an experiment with the product initial state and observes Bell violation at some scattering scale, then the constraints in $\text{eq}^\text{n}$(\ref{QED_con}) must hold true.
However, we do not find any clear \textit{theoretical} motivation to demand Bell violation in $2\rightarrow2$ scattering with product state as the initial state, as explored by \cite{sinha_10.48550}. In fact, on the contrary, we will show that for a product initial state, there is no Bell violation at any scattering angle for EFT of gluons (non-abelian), with a weakly coupled UV completion.

\subsection{EFT for gluons}\label{non-abelian}
Now let's consider the following lagrangian containing only the CP conserving operator for EFT of gluons upto dim 6,
\begin{equation}
    L=-\frac{1}{4}G^a_{\mu\nu}G^{a\mu\nu}+g^3\frac{c_1}{\Lambda^2}f^{abc} G_\mu^{a\nu}G_\nu^{b\rho}G_\rho^{c\mu}
\end{equation}
For the above Lagrangian, we get the following helicity amplitudes up to $\mathcal{O}(1/\Lambda^2)$ for process $|p_1,\epsilon_1,a;p_2,\epsilon_2,b \rangle  \rightarrow |p_3,\epsilon_3,d;p_4,\epsilon_4,e \rangle$ 
%
%
\begin{align}
& \mathcal{M}_{1,1}^{1,1}(s, t, u,a,b,d,e)=2g^2 \frac{s}{tu}(f^{acb}f^{dce}u-sf^{ace}f^{bcd}) \\\nn
& \mathcal{M}_{1,1}^{0,0}(s, t, u,a,b,d,e)=-12g^4\frac{c_1}{\Lambda^2}f^{acb}f^{dce}\frac{(t^2+u^2)}{(t-u)}-12g^4\frac{c_1s}{\Lambda^2}\frac{(tf^{acd}f^{bce}-uf^{ace}f^{bcd})}{(t-u)}\\\nn
& \mathcal{M}_{1,0}^{1,0}(s, t, u,a,b,d,e)=2g^2 \frac{u}{ts}(f^{ace}f^{dcb}s-uf^{acb}f^{ecd}) \\\nn
& \mathcal{M}_{1,0}^{0,1}(s, t, u,a,b,d,e)=2g^2 \frac{t}{us}(f^{acd}f^{bce}u-tf^{ace}f^{dcb}) \\\nn
& \mathcal{M}_{1,1}^{1,0}(s, t, u,a,b,d,e)=6g^4\frac{c_1}{\Lambda^2}(uf^{acd}f^{bce}+tf^{ace}f^{bcd})
\end{align}
where $a$, $b$, $c$, $d$, and $e$ represent the color of particles and $c$ is summed over.

We sum over the colors of final states, $d$ and $e$, which makes the final state entangled only in the helicity basis. Then the amplitudes reduce to the following forms,
\begin{align}
& \Phi_1(s, t, u)=-2g^2 \frac{s^2}{tu}\left(\sum_{d,e}f^{ace}f^{bcd}\right) \quad;\quad \Phi_2(s, t, u)=-12g^4\frac{c_1s}{\Lambda^2}\left(\sum_{d,e}f^{ace}f^{bcd}\right)\\\nn
&\Phi_3(s, t, u)=-2g^2 \frac{u}{t}\left(\sum_{d,e}f^{ace}f^{bcd}\right) \quad;\quad
\Phi_4(s, t, u)=-2g^2 \frac{t}{u}\left(\sum_{d,e}f^{ace}f^{bcd}\right) \\\nn
& \Phi_5(s, t, u)=-6g^4\frac{c_1s}{\Lambda^2}\left(\sum_{d,e}f^{ace}f^{bcd}\right)
\end{align}
For the generic initial state, $\ket{\psi}_i=\cos\theta\ket{0,0}+\sin\theta\ket{1,1}$, we get the following CGLMP Bell parameter corresponding to the final state (details of which have been relegated to the appendix),
\begin{equation}\label{I_2f}
     I_{2f}=\frac{2\sqrt{2}\l(2\Phi_1\Phi_2+(\Phi_1^2+\Phi_2^2)\sin2\theta-2\Phi_5^2(1+\sin2\theta)\r)}{|\Phi_1|^2+|\Phi_2|^2+2\Phi_1\Phi_2\sin2\theta+2\Phi_5^2(1+\sin2\theta)}
\end{equation}
For the above $\Phi_1$, $\Phi_2$ and $\Phi_5$ we get the following $I_{2f}$,
\begin{align}
    I_{2f}&=2\sqrt{2}\;\frac{48c_1'uts^2+(4s^4+144c_1'^2u^2t^2)\sin2\theta-72c_1'^2u^2t^2(1+\sin2\theta)}{48c_1's^2ut\sin2\theta+4s^4+144c_1'^2u^2t^2+72c_1'^2u^2t^2(1+\sin2\theta)}\\\nn
    &=2\sqrt{2}\;\frac{\sin2\theta-12c_1'\chi(1+\chi)+18c_1'^2\chi^2(1+\chi)^2(\sin2\theta-1)}{1-12c_1'\chi(1+\chi)\sin2\theta+18c_1'^2\chi^2(1+\chi)^2(\sin2\theta+3)}
\end{align}
where $c_1'=g^2c_1s/\Lambda^2$.
We observe Bell violation for some $\chi$ and $\theta$ (with the constraint $|I_{2i}|\leq2$) given 
\begin{equation}\label{c_1'}
    c_1'\in\l(-\infty,-\frac{2\sqrt{2}}{3(3\sqrt{2}-4)}\r)\cup\l(-\frac{2\sqrt{2}}{3(3\sqrt{2}+4)},\infty\r)\setminus\{0\}
\end{equation}
Since $c_1'=g^2c_1s/\Lambda^2$ and $s<\Lambda^2$ within the validity of the EFT regime, we finally get
\begin{equation}
    c_1\in\mathbb{R}\setminus{0}
\end{equation}
For any $c_1$ except $0$, we can choose appropriate $s$ so that $c_1'$ lies in the range (\ref{c_1'}).Therefore, for all values of $c_1$ except $0$, Bell inequalities can be violated at some scattering angle for some initial state with $|I_{2i}|\leq2$.
\vspace{2mm}

\noindent\textbf{Product initial state:} Now, if we take the product state $\ket{1,1}$ to be the initial state, we get the following $I_{2f}$,
\begin{equation}
    I_{2f}=-2\sqrt{2}\frac{12c_1'\chi(1+\chi)+18c_1'^2\chi^2(1+\chi)^2}{1+54c_1'^2\chi^2(1+\chi)^2}
\end{equation}
We observe Bell violation for some $\chi$ if,
\begin{equation}
    c_1'<\frac{-4\sqrt{2}+2\sqrt{2(1+\sqrt{2})}}{3(3-\sqrt{2})}\approx-0.265
\end{equation}
and since $s<\Lambda^2$ and $g\sim\mathcal{O}(1)$, $c_1$ has to be at least of $\mathcal{O}(1)$. However, the value of $c_1$ is expected to be of much smaller order for weakly coupled theories; for example, for weakly coupled UV completion with heavy fermions, one typically gets $c_1$ of $\mathcal{O}(10^{-4})$ as shown in \cite{1810.06994}. Thus, we don't expect to observe Bell violation by the non-abelian gauge theory in $2\rightarrow2$ scattering for the product state as the initial state.

It is interesting that this is qualitatively different from the abelian case of QED, where it is possible to observe Bell violation for the product initial state. This qualitative difference between abelian and non-abelian gauge theory is mainly due to the contribution from the kinetic term of non-abelian gauge theory towards the MHV amplitude. This contribution doesn't allow the cancellation of energy scale $\Lambda$ in $I_{2f}$.

This also shows that Bell violation (for product initial state) cannot be promoted to a principle to constrain EFTs and the QED value satisfying the constraint (\ref{QED_con}) is perhaps just a coincidence.


\subsection{Bell inequality for $RF^2$}
Now we consider $2\rightarrow2$ scattering of photons, including the graviton exchange.
We use the following curvature conventions for the calculations,
\begin{align} 
\Gamma_{\mu \nu}^\lambda & =\frac{1}{2} g^{\lambda \sigma}\left[\partial_\mu g_{\nu \sigma}+\partial_\nu g_{\sigma \mu}-\partial_\sigma g_{\mu \nu}\right] \\ R_{\sigma \mu \nu}^\rho & =\partial_\mu \Gamma_{\nu \sigma}^\rho-\partial_\nu \Gamma_{\mu \sigma}^\rho+\Gamma_{\mu \lambda}^\rho \Gamma_{\nu \sigma}^\lambda-\Gamma_{\nu \lambda}^\rho \Gamma_{\mu \sigma}^\lambda, \quad R_{\sigma \nu}=R_{\sigma \rho \nu}^\rho,
\end{align}
Consider the following EFT action for photon coupled to gravity,
\begin{equation}
S=\int \mathrm{d}^4 x \sqrt{-g}\left[-\frac{2}{\kappa^2} R-\frac{1}{4} F_{\mu \nu}^2+\frac{\Hat{\alpha}}{4\Lambda^2} R_{\mu \nu \rho \sigma} F^{\mu \nu} F^{\rho \sigma}+\frac{c_1}{\Lambda^4}(F^{\mu\nu}F_{\mu\nu})^2+\frac{c_2}{\Lambda^4}(F^{\mu\nu}\widetilde{F}_{\mu\nu})^2\right]
\end{equation}
where $\kappa=2/M_{\text{pl}}$. Taking the gravity to be perturbative i.e. $g_{\mu\nu}=\eta_{\mu\nu}+\kappa h_{\mu\nu}$, we get the following helicity amplitudes at tree level,
\begin{align}
    \mathcal{M}_{1,1}^{1,1}(s,t,u)&=\Phi_1=g_2s^2+\frac{s\kappa^2}{16}\left[\left(\alpha^2+\frac{4}{ut}\right)\left(s^2+ut\right)+6\alpha s\right]\\\nn
    \mathcal{M}_{1,1}^{0,0}(s,t,u)&=\Phi_2=f_2(s^2+t^2+u^2)-\frac{\alpha\kappa^2}{16}\left[2\left(s^2+t^2+u^2\right)-3\alpha stu\right]\\\nn
    \mathcal{M}_{1,0}^{1,0}(s,t,u)&=\Phi_5=-\frac{\alpha\kappa^2}{16}\left[s^2+t^2+u^2+3\alpha stu\right]
\end{align}
where $f_2$ and $g_2$ are same as defined in $\text{eq}^\text{n}$(\ref{photon_amp}) and $\alpha=\Hat{\alpha}/\Lambda^2$.
\vspace{2mm}

 \noindent For the spinor QED UV completion, $\Lambda=m_e$ (mass of electron) and for $s\ll m_e^2$ we have \cite{2012.05798}
\begin{equation}
    f_2=\frac{-e^4}{240\pi^2m_e^4} \quad;\quad g_2=\frac{11e^4}{720\pi^2m_e^4}\quad;\quad \alpha=\frac{-e^2}{360\pi^2m_e^2}
\end{equation}
The $I_{2f}$ is same as $\text{eq}^\text{n}$(\ref{I_2f}) for a generic initial state: $\cos\theta\ket{0,0}+\sin\theta\ket{1,1}$,
\begin{equation}
     I_{2f}=\frac{2\sqrt{2}\l(2\Phi_1\Phi_2+(\Phi_1^2+\Phi_2^2)\sin2\theta-2\Phi_5^2(1+\sin2\theta)\r)}{|\Phi_1|^2+|\Phi_2|^2+2\Phi_1\Phi_2\sin2\theta+2\Phi_5^2(1+\sin2\theta)}
\end{equation}
%
After calculating $I_{2f}$ for the above amplitudes and Wilson coefficients, we observe Bell violation for all values of $(e^2M_{pl}^2)/m_e^2$ for some scattering angle and $\theta$.
\vspace{2mm}

\noindent \textbf{Product initial state:} Taking the product state as the initial state, we observe Bell violation if,
\begin{equation}\label{WGC_con}
    \frac{e M_{pl}}{m_e}\geq f(s/M_\text{pl})
\end{equation}
where $f$ is some function of $s/M_\text{pl}$ which increases with decreasing $s/M_\text{pl}$. For example, $f(s/M_{pl})\sim195$ for $s = M_{pl}^2$ and it increases to $f(s/M_{pl})\sim1946$ for $s=0.01  M_{pl}^2$.

The above constraint (\ref{WGC_con}) is similar to the \textit{Weak Gravity Conjecture} (WGC) ($e/m\geq\mathcal{O}(1)/M_{\text{pl}}$) \cite{WGC} for $f\sim\mathcal{O}(1)$, as noted by \cite{sinha_10.48550}. However, within the validity of the EFT regime, $ s<\Lambda^2\ll M_{\text{pl}}$, $f$ is of a much higher order than $\mathcal{O}(1)$ and therefore cannot be compared to WGC. If we experimentally observe Bell violation for product initial state, then we get much stronger constraints on the charge-to-mass ratio of fermions coupled to photons than imposed by WGC.


\subsection{Gravity}
Consider the following action for gravity, including the corrections to Einstein's gravity
\begin{equation}
    S=\int d^4x\sqrt{-g} \frac{2}{\kappa^2}\l(R+\frac{\Hat{\beta}}{3!\Lambda^4}R^3\r)
\end{equation}
where $R^3\equiv R^{\mu\nu\kappa\lambda}R_{\kappa\lambda\alpha\gamma}R^{\alpha\gamma}_{\;\;\;\;\;\mu\nu}$.

Again taking gravity to be perturbative $g_{\mu\nu}=\eta_{\mu\nu}+\kappa h_{\mu\nu}$, we get the following helicity amplitudes for $2\rightarrow2$ scattering of gravitons \cite{2103.12728}
\begin{align}
    \mathcal{M}_{1,1}^{1,1}(s,t,u)&=\Phi_1=\kappa^2s\l(\frac{s^2}{ut}+\frac{\beta^2}{16}s^2ut\r)\\\nn
    \mathcal{M}_{1,1}^{0,0}(s,t,u)&=\Phi_2=\frac{5}{4}\kappa^2\beta stu\\\nn
    \mathcal{M}_{1,0}^{1,0}(s,t,u)&=\Phi_5=\frac{1}{8}\kappa^2\beta stu
\end{align}
where $\beta=\Hat{\beta}/\Lambda^4$.
For a generic initial state, $\cos\theta\ket{0,0}+\sin\theta\ket{1,1}$, we have the following form of $I_{2f}$
\begin{equation}
     I_{2f}=\frac{2\sqrt{2}\l(2\Phi_1\Phi_2+(\Phi_1^2+\Phi_2^2)\sin2\theta-2\Phi_5^2(1+\sin2\theta)\r)}{|\Phi_1|^2+|\Phi_2|^2+2\Phi_1\Phi_2\sin2\theta+2\Phi_5^2(1+\sin2\theta)}
\end{equation}
In this case, as well, we observe Bell violation, i.e., $|I_{2f}|>2$ for all values of the Wilson coefficient $\Hat{\beta}$ for some $\chi$ and $\theta$ (with the constraint $|I_{2i}|\leq2$).
\vspace{2mm}

\noindent \textbf{Product initial state:} For the product initial state, we observe Bell violation at some scattering angle if $1.379<|\beta| s^2<46.417$. Since $s<\Lambda^2$ within the validity of EFT regime, $\Hat{\beta}$ must be of a much higher order than $\mathcal{O}(1)$. If one considers $s\sim\Lambda^2$, then $\Hat{\beta}$ must be of $\mathcal{O}(1)$ to observe Bell violation; however, then one has to consider higher-dimensional operators as their contributions also become significant. This is similar to the case of non-abelian gauge theory in section (\ref{non-abelian}).


\section{CP conserving vs CP violating}\label{CP_vs_CPv}
In this section, we are not trying to probe the quantum nature of theory but to see if we can differentiate between CP-conserving and CP-violating theories using entanglement and $2\rightarrow2$ scattering. The CP-violating terms in the Lagrangian of a theory give imaginary contributions to the helicity amplitudes which lead to complex coefficients in the final state. Since the CGLMP Bell parameter is not a good measure of entanglement in states with complex coefficients, we will use another parameter, \textit{concurrence} ($\Delta$), for this purpose. Concurrence is defined as 
$\Delta=2\;|\mu_{00}\mu_{11}-\mu_{01}\mu_{10}|$
for a normalized state, $\ket{\Psi}=\mu_{mn}\ket{m,n}$. The $\Delta=1$ corresponds to a maximally entangled state, whereas $\Delta=0$ corresponds to a product state.

We consider $\theta=-\pi/4$ in the initial state $\ket{\psi}_i=\cos\theta\ket{0,0}+\sin\theta\ket{1,1}$ i.e. it is maximally entangled.
We have the following helicity amplitudes after including CP-violating terms in the theory,
$$
\left(\begin{array}{cccc}
\mathcal{M}_{1,0}^{1,0} & \mathcal{M}_{1,0}^{1,1} & \mathcal{M}_{0,0}^{1,0} & \mathcal{M}_{0,0}^{1,1} \\
\mathcal{M}_{1,0}^{0,0} & \mathcal{M}_{1,0}^{0,1} & \mathcal{M}_{0,0}^{0,0} & \mathcal{M}_{0,0}^{0,1} \\
\mathcal{M}_{1,1}^{1,0} & \mathcal{M}_{1,1}^{1,1} & \mathcal{M}_{0,1}^{1,0} & \mathcal{M}_{0,1}^{1,1} \\
\mathcal{M}_{1,1}^{0,0} & \mathcal{M}_{1,1}^{0,1} & \mathcal{M}_{0,1}^{0,0} & \mathcal{M}_{0,1}^{0,1}
\end{array}\right)=\left(\begin{array}{cccc}
\Phi_3 & \Phi_5^* & \Phi_5^* & \Phi_2^* \\
\Phi_5 & \Phi_4 & \Phi_1^* & \Phi_5^* \\
\Phi_5 & \Phi_1 & \Phi_4 & \Phi_5^* \\
\Phi_2 & \Phi_5 & \Phi_5 & \Phi_3
\end{array}\right)
$$
Then we get the following final scattered state for the maximally entangled initial state,
\begin{align}
      \ket{\Psi}=&(\Phi_1^*-\Phi_2)\ket{0,0}+(\Phi_2^*-\Phi_1)\ket{1,1}+\l(\Phi_5^*-\Phi_5\r)\l(\ket{0,1}+\ket{1,0}\r)
  \end{align}
which has the concurrence ($\Delta_f$),
\begin{equation}
    \Delta_f=\frac{\l||\Phi_1^*-\Phi_2|^2+(\Phi_5-\Phi_5^*)^2\r|}{|\Phi_1^*-\Phi_2|^2+|\Phi_5-\Phi_5^*|^2}=\frac{\l||\Phi_1^*-\Phi_2|^2-4(\text{Im}\Phi_5)^2\r|}{|\Phi_1^*-\Phi_2|^2+4(\text{Im}\Phi_5)^2}
\end{equation}
In the case of CP conserving theories, we have real amplitudes; therefore, $\Phi_5=\Phi_5^*$ and $\Delta=1$, i.e. the final state is maximally entangled. However, in general, we can have $\Delta<1$ for the CP-violating theories.
Therefore, if one observes a non-maximally entangled final state ($\Delta<1$) in $2\rightarrow2$ scattering with $\ket{\psi}_i=(\ket{0,0}-\ket{1,1})/\sqrt{2}$ as the initial state, then the theory has CP-violating contributions. Note that this statement is true about the full theory since we have used general amplitudes which are not limited to a certain order in $\Lambda$.

\section{Discussion}\label{summary}
We explored Bell violation for $2\rightarrow2$ scattering of photons, gluons and gravitons in the context of EFTs using the CGLMP Bell parameter as the measure of entanglement. 
We considered the initial state to be entangled in the Hilbert space spanned by the helicity basis, such that the degree of entanglement can be described by a local hidden-variable theory. This condition on the initial state can be described as the relation $|I_{2i}|\leq2$, where the $I_{2i}$ represents the CGLMP parameter. With this particular choice of the initial state, the Bell inequality for the final state can be violated only due to the quantum nature of the scattering amplitudes, which is dictated by the theory in consideration. 
\vspace{2mm}

\noindent We showed that starting from an appropriate initial state, $2\rightarrow2$ scattering of photons, gluons, and gravitons could violate the Bell inequality (at least for some scattering angle) for any non-zero value of CP-conserving higher dimensional operators in the corresponding EFTs.
\vspace{2mm}

\noindent If one considers the initial state to be a product state, which is experimentally easier to prepare, and observes Bell violation at some scattering angle, then the EFT parameter space can be constrained. This was also shown by \cite{sinha_10.48550} for the QED case. However, a priori, one can not use this as a principle to constrain the EFTs as we have explicitly shown using the example of the EFT of gluons. In the cases of EFTs for gluons, gravity, and photons including gravity, we observe Bell violation for the initial product state (say $\ket{1,1}$) if the Wilson coefficients of higher dimensional operators are of at least $\mathcal{O}(1)$. In all these cases, the leading operator (4-dim operator) contributes only to the MHV amplitude ($\Phi_1=\mathcal{M}_{1,1}^{1,1}$), therefore one basis ($\ket{1,1}$) has a significantly higher weight than the other ($\ket{0,0}$) (even after Schmidt decomposition). However, for a significant degree of entanglement in the final state ($I_{2f}>2$), we need the weights of both bases to be comparable. Thus, the Wilson coefficient must be at least $\mathcal{O}(1)$, so that both the bases have comparable weights. It is interesting that the non-abelian gauge theory (and gravity) is qualitatively different from the abelian gauge theory even from the point of view of Bell violation.
\vspace{2mm}

\noindent We have also shown that if we consider the initial state in $2\rightarrow2$ scattering to be a particular maximally entangled state, then we can probe the CP-violating nature of the theory using the degree of entanglement in final states.
\vspace{2mm}

In this work, we have explored the possibility of Bell violation by the unitary evolution of qubits for different EFTs using $2\rightarrow2$ scattering. There is still much to explore on the relationship between entanglement and EFTs. It would be interesting to explore Bell violation by states entangled with respect to quantum numbers other than helicity, like colors for gluons, and if it can restrict the EFT parameter space. It would also be interesting to investigate whether our results hold true even after considering more higher-dimensional operators in the EFT, like dim 8 operators in the EFT of gluons.

\section*{Acknowledgments}
DG acknowledges support through the Ramanujan Fellowship
and MATRICS Grant of the Department of Science and Technology, Government of India. The authors thank Farman
Ullah for discussions, and Arijit Chatterjee for discussions and comments on the draft.

\appendix  
\section*{Appendix}
    \section{Schmidt decomposition}
  For a generic initial state, $\ket{\psi}_i=\cos\theta\ket{0,0}+\sin\theta\ket{1,1}$, we get the following final state after $2\rightarrow2$ scattering
  \begin{align}
      \ket{\Psi}=&\cos\theta(\mathcal{M}_{0,0}^{0,0}\ket{0,0}+\mathcal{M}_{0,0}^{1,1}\ket{1,1}+\mathcal{M}_{0,0}^{0,1}\ket{0,1}+\mathcal{M}_{0,0}^{1,0}\ket{1,0})\\\nn
      &\sin\theta(\mathcal{M}_{1,1}^{0,0}\ket{0,0}+\mathcal{M}_{1,1}^{1,1}\ket{1,1}+\mathcal{M}_{1,1}^{0,1}\ket{0,1}+\mathcal{M}_{1,1}^{1,0}\ket{1,0})
  \end{align}
  For CP conserving theories, we have
  $$
\left(\begin{array}{cccc}
\mathcal{M}_{1,0}^{1,0} & \mathcal{M}_{1,0}^{1,1} & \mathcal{M}_{0,0}^{1,0} & \mathcal{M}_{0,0}^{1,1} \\
\mathcal{M}_{1,0}^{0,0} & \mathcal{M}_{1,0}^{0,1} & \mathcal{M}_{0,0}^{0,0} & \mathcal{M}_{0,0}^{0,1} \\
\mathcal{M}_{1,1}^{1,0} & \mathcal{M}_{1,1}^{1,1} & \mathcal{M}_{0,1}^{1,0} & \mathcal{M}_{0,1}^{1,1} \\
\mathcal{M}_{1,1}^{0,0} & \mathcal{M}_{1,1}^{0,1} & \mathcal{M}_{0,1}^{0,0} & \mathcal{M}_{0,1}^{0,1}
\end{array}\right)=\left(\begin{array}{cccc}
\Phi_3 & \Phi_5 & \Phi_5 & \Phi_2 \\
\Phi_5 & \Phi_4 & \Phi_1 & \Phi_5 \\
\Phi_5 & \Phi_1 & \Phi_4 & \Phi_5 \\
\Phi_2 & \Phi_5 & \Phi_5 & \Phi_3
\end{array}\right)
$$
Therefore, the normalized final state can be written as
  \begin{align}
      \ket{\Psi}&=\frac{(\sin\theta\Phi_2+\cos\theta\Phi_1)\ket{0,0}+(\sin\theta\Phi_1+\cos\theta\Phi_2)\ket{1,1}+(\sin\theta+\cos\theta)\Phi_5(\ket{0,1}+\ket{1,0})}{\sqrt{|\Phi_1|^2+|\Phi_2|^2+2\Phi_1\Phi_2\sin2\theta+2\Phi_5^2(1+\sin2\theta)}}\\\nn
      &=\mu_{mn}\ket{m,n}
  \end{align}
  Now, we will use the Schmidt decomposition theorem \cite{schmidt} to convert the above state to $\ket{0,0}$ and $\ket{1,1}$ basis.
The coefficients $\mu_{mn}$ can be written in the form of a matrix,
$$\mu_{mn}=\l(\begin{array}{cc}
   (\sin\theta\Phi_2+\cos\theta\Phi_1)  & (\sin\theta+\cos\theta)\Phi_5 \\
    (\sin\theta+\cos\theta)\Phi_5 & (\sin\theta\Phi_1+\cos\theta\Phi_2)
\end{array}\r)/\sqrt{|\Phi_1|^2+|\Phi_2|^2+2\Phi_1\Phi_2\sin2\theta+2\Phi_5^2(1+\sin2\theta)}
$$
We then diagonalize the above matrix, i.e. find the eigenvalues of the matrix, which are given by the roots of the following equation
\begin{align}
    \lambda^2&-\frac{(\Phi_1+\Phi_2)(\sin\theta+\cos\theta)\lambda}{\sqrt{|\Phi_1|^2+|\Phi_2|^2+2\Phi_1\Phi_2\sin2\theta+2\Phi_5^2(1+\sin2\theta)}}\\\nn
    &+\frac{\Phi_1\Phi_2+(\Phi_1^2+\Phi_2^2)\sin\theta\cos\theta-\Phi_5^2(1+\sin2\theta)}{(|\Phi_1|^2+|\Phi_2|^2+2\Phi_1\Phi_2\sin2\theta+2\Phi_5^2(1+\sin2\theta))}=0
\end{align}
These eigenvalues are the new coefficients in $\ket{00}$ and $\ket{11}$ basis.
We don't need to explicitly calculate the roots of the above equation as the CGLMP depends just on their product,
\begin{align}
    I_{2f}&=4\sqrt{2}\lambda_1\lambda_2\\\nn
    &=2\sqrt{2}\frac{2\Phi_1\Phi_2+(\Phi_1^2+\Phi_2^2)\sin2\theta-2
    \Phi_5^2(1+\sin2\theta)}{|\Phi_1|^2+|\Phi_2|^2+2\Phi_1\Phi_2\sin2\theta+2\Phi_5^2(1+\sin2\theta)}
\end{align}
One can easily see that in terms of $\mu_{mn}$, we have $\lambda_1\lambda_2=\mu_{00}\mu_{11}-\mu_{01}\mu_{10}$. Therefore, $I_{2}$ can now be directly written as
\begin{equation}
    I_{2f}=4\sqrt{2}\frac{\mu_{00}\mu_{11}-\mu_{01}\mu_{10}}{\sum_{m,n=0}^1|\mu_{m,n}|^2}
\end{equation}
for a generic state, $\ket{\Psi}=\sum_{m,n=0}^1\mu_{mn}\ket{m,n}$ and real $\mu_{mn}$.
\vspace{2mm}

\noindent The above result could have also been inferred by directly looking at another parameter quantifying the degree of entanglement, \textit{concurrence} ($\Delta$), which is defined as 
$\Delta=2\;|\mu_{00}\mu_{11}-\mu_{01}\mu_{10}|$
for a normalized state. This reduces to $\mu'_{00}\mu'_{11}$ when one goes to $\ket{00}$ and $\ket{11}$ basis as $\mu'_{01}=0=\mu'_{10}$. Since the degree of entanglement doesn't depend on the basis, we can infer the relation between the coefficients in different basis as $\mu_{00}\mu_{11}-\mu_{01}\mu_{10}=\mu'_{00}\mu'_{11}$ which is same as we derived from Schmidt decomposition method.

\nocite{*}
\color{black}
\bibliographystyle{JHEP}
{\footnotesize
\bibliography{ref}}

\providecommand{\href}[2]{#2}\begingroup\raggedright\begin{thebibliography}{10}

\bibitem{Bell}
J.~S. Bell, {\it On the einstein podolsky rosen paradox},  {\em Physics
  Physique Fizika} {\bf 1} (Nov, 1964) 195--200.

\bibitem{PhysRevLett.115.250401}
M.~Giustina, M.~A.~M. Versteegh, S.~Wengerowsky, J.~Handsteiner, A.~Hochrainer,
  K.~Phelan, F.~Steinlechner, J.~Kofler, J.-A. Larsson, C.~Abell\'an, W.~Amaya,
  V.~Pruneri, M.~W. Mitchell, J.~Beyer, T.~Gerrits, A.~E. Lita, L.~K. Shalm,
  S.~W. Nam, T.~Scheidl, R.~Ursin, B.~Wittmann, and A.~Zeilinger, {\it
  Significant-loophole-free test of bell's theorem with entangled photons},
  {\em Phys. Rev. Lett.} {\bf 115} (Dec, 2015) 250401.

\bibitem{PhysRevLett.28.938}
S.~J. Freedman and J.~F. Clauser, {\it Experimental test of local
  hidden-variable theories},  {\em Phys. Rev. Lett.} {\bf 28} (Apr, 1972)
  938--941.

\bibitem{PhysRevLett.49.1804}
A.~Aspect, J.~Dalibard, and G.~Roger, {\it Experimental test of bell's
  inequalities using time-varying analyzers},  {\em Phys. Rev. Lett.} {\bf 49}
  (Dec, 1982) 1804--1807.

\bibitem{PhysRevLett.115.250402}
L.~K. Shalm, E.~Meyer-Scott, Christensen, and et~al, {\it Strong loophole-free
  test of local realism}, .

\bibitem{1206.2031}
W.~{Pfaff}, T.~H. {Taminiau}, L.~{Robledo}, H.~{Bernien}, M.~{Markham}, D.~J.
  {Twitchen}, and R.~{Hanson}, {\it {Demonstration of
  entanglement-by-measurement of solid-state qubits}},  {\em Nature Physics}
  {\bf 9} (Jan., 2013) 29--33, [\href{http://arxiv.org/abs/1206.2031}{{\tt
  arXiv:1206.2031}}].

\bibitem{1508.05949}
B.~{Hensen}, H.~{Bernien}, A.~E. {Dr{\'e}au}, A.~{Reiserer}, N.~{Kalb}, M.~S.
  {Blok}, J.~{Ruitenberg}, R.~F.~L. {Vermeulen}, R.~N. {Schouten},
  C.~{Abell{\'a}n}, W.~{Amaya}, V.~{Pruneri}, M.~W. {Mitchell}, M.~{Markham},
  D.~J. {Twitchen}, D.~{Elkouss}, S.~{Wehner}, T.~H. {Taminiau}, and
  R.~{Hanson}, {\it {Loophole-free Bell inequality violation using electron
  spins separated by 1.3 kilometres}},  {\em Nature} {\bf 526} (Oct., 2015)
  682--686, [\href{http://arxiv.org/abs/1508.05949}{{\tt arXiv:1508.05949}}].

\bibitem{2201.13310}
H.~{Casini} and M.~{Huerta}, {\it {Lectures on entanglement in quantum field
  theory}},  {\em arXiv e-prints} (Jan., 2022) arXiv:2201.13310,
  [\href{http://arxiv.org/abs/2201.13310}{{\tt arXiv:2201.13310}}].

\bibitem{1803.04993}
E.~{Witten}, {\it {Notes on Some Entanglement Properties of Quantum Field
  Theory}},  {\em arXiv e-prints} (Mar., 2018) arXiv:1803.04993,
  [\href{http://arxiv.org/abs/1803.04993}{{\tt arXiv:1803.04993}}].

\bibitem{0912.1941}
M.~{Junge}, C.~{Palazuelos}, D.~{P{\'e}rez-Garc{\'\i}a}, I.~{Villanueva}, and
  M.~M. {Wolf}, {\it {Operator Space Theory: A Natural Framework for Bell
  Inequalities}},  {\em PhysRevLett} {\bf 104} (Apr., 2010) 170405,
  [\href{http://arxiv.org/abs/0912.1941}{{\tt arXiv:0912.1941}}].

\bibitem{2003.02280}
Y.~{Afik} and J.~{Ram{\'o}n Mu{\~n}oz de Nova}, {\it {Entanglement and quantum
  tomography with top quarks at the LHC}},  {\em arXiv e-prints} (Mar., 2020)
  arXiv:2003.02280, [\href{http://arxiv.org/abs/2003.02280}{{\tt
  arXiv:2003.02280}}].

\bibitem{2102.11883}
M.~{Fabbrichesi}, R.~{Floreanini}, and G.~{Panizzo}, {\it {Testing Bell
  Inequalities at the LHC with Top-Quark Pairs}},  {\em PhysRevLett.} {\bf 127}
  (Oct., 2021) 161801, [\href{http://arxiv.org/abs/2102.11883}{{\tt
  arXiv:2102.11883}}].

\bibitem{2110.10112}
C.~{Severi}, C.~D.~E. {Boschi}, F.~{Maltoni}, and M.~{Sioli}, {\it {Quantum
  tops at the LHC: from entanglement to Bell inequalities}},  {\em European
  Physical Journal C} {\bf 82} (Apr., 2022) 285,
  [\href{http://arxiv.org/abs/2110.10112}{{\tt arXiv:2110.10112}}].

\bibitem{2205.00542}
J.~A. {Aguilar-Saavedra} and J.~A. {Casas}, {\it {Improved tests of
  entanglement and Bell inequalities with LHC tops}},  {\em European Physical
  Journal C} {\bf 82} (Aug., 2022) 666,
  [\href{http://arxiv.org/abs/2205.00542}{{\tt arXiv:2205.00542}}].

\bibitem{2203.05582}
Y.~{Afik} and J.~R. M.~d. {Nova}, {\it {Quantum information with top quarks in
  QCD}},  {\em Quantum} {\bf 6} (Sept., 2022) 820,
  [\href{http://arxiv.org/abs/2203.05582}{{\tt arXiv:2203.05582}}].

\bibitem{2107.13007}
W.~{Gong}, G.~{Parida}, Z.~{Tu}, and R.~{Venugopalan}, {\it {Measurement of
  Bell-type inequalities and quantum entanglement from {\ensuremath{\Lambda}}
  -hyperon spin correlations at high energy colliders}},  {\em PhysRevD} {\bf
  106} (Aug., 2022) L031501, [\href{http://arxiv.org/abs/2107.13007}{{\tt
  arXiv:2107.13007}}].

\bibitem{2106.01377}
A.~J. {Barr}, {\it {Testing Bell inequalities in Higgs boson decays}},  {\em
  Physics Letters B} {\bf 825} (Feb., 2022) 136866,
  [\href{http://arxiv.org/abs/2106.01377}{{\tt arXiv:2106.01377}}].

\bibitem{2204.11063}
A.~J. {Barr}, P.~{Caban}, and J.~{Rembieli{\'n}ski}, {\it {Bell-type
  inequalities for systems of relativistic vector bosons}},  {\em arXiv
  e-prints} (Apr., 2022) arXiv:2204.11063,
  [\href{http://arxiv.org/abs/2204.11063}{{\tt arXiv:2204.11063}}].

\bibitem{1703.02989}
A.~{Cervera-Lierta}, J.~{Latorre}, J.~{Rojo}, and L.~{Rottoli}, {\it {Maximal
  Entanglement in High Energy Physics}},  {\em SciPost Physics} {\bf 3} (Nov.,
  2017) 036, [\href{http://arxiv.org/abs/1703.02989}{{\tt arXiv:1703.02989}}].

\bibitem{2108.00646}
S.~R. {Beane}, R.~C. {Farrell}, and M.~{Varma}, {\it {Entanglement minimization
  in hadronic scattering with pions}},  {\em International Journal of Modern
  Physics A} {\bf 36} (Oct., 2021) 2150205--214,
  [\href{http://arxiv.org/abs/2108.00646}{{\tt arXiv:2108.00646}}].

\bibitem{1812.03138}
S.~R. {Beane}, D.~B. {Kaplan}, N.~{Klco}, and M.~J. {Savage}, {\it
  {Entanglement Suppression and Emergent Symmetries of Strong Interactions}},
  {\em PhysRevLett} {\bf 122} (Mar., 2019) 102001,
  [\href{http://arxiv.org/abs/1812.03138}{{\tt arXiv:1812.03138}}].

\bibitem{2208.11723}
M.~{Fabbrichesi}, R.~{Floreanini}, and E.~{Gabrielli}, {\it {Constraining new
  physics in entangled two-qubit systems: top-quark, tau-lepton and photon
  pairs}},  {\em arXiv e-prints} (Aug., 2022) arXiv:2208.11723,
  [\href{http://arxiv.org/abs/2208.11723}{{\tt arXiv:2208.11723}}].

\bibitem{2203.05619}
R.~{Aoude}, E.~{Madge}, F.~{Maltoni}, and L.~{Mantani}, {\it {Quantum SMEFT
  tomography: Top quark pair production at the LHC}},  {\em PhysRevD} {\bf 106}
  (Sept., 2022) 055007, [\href{http://arxiv.org/abs/2203.05619}{{\tt
  arXiv:2203.05619}}].

\bibitem{sinha_10.48550}
A.~{Sinha} and A.~{Zahed}, {\it {Bell inequalities in 2-2 scattering}},  {\em
  arXiv e-prints} (Dec., 2022) arXiv:2212.10213,
  [\href{http://arxiv.org/abs/2212.10213}{{\tt arXiv:2212.10213}}].

\bibitem{2210.09330}
C.~{Severi} and E.~{Vryonidou}, {\it {Quantum entanglement and top spin
  correlations in SMEFT at higher orders}},  {\em Journal of High Energy
  Physics} {\bf 2023} (Jan., 2023) 148,
  [\href{http://arxiv.org/abs/2210.09330}{{\tt arXiv:2210.09330}}].

\bibitem{040404}
D.~{Collins}, N.~{Gisin}, N.~{Linden}, S.~{Massar}, and S.~{Popescu}, {\it
  {Bell Inequalities for Arbitrarily High-Dimensional Systems}},  {\em Phys.
  Rev. Lett.} {\bf 88} (Jan., 2002) 040404,
  [\href{http://arxiv.org/abs/quant-ph/0106024}{{\tt quant-ph/0106024}}].

\bibitem{CHSH}
J.~F. Clauser, M.~A. Horne, A.~Shimony, and R.~A. Holt, {\it Proposed
  experiment to test local hidden-variable theories},  {\em Phys. Rev. Lett.}
  {\bf 23} (Oct, 1969) 880--884.

\bibitem{0101084}
T.~{Durt}, D.~{Kaszlikowski}, and M.~{{\.Z}ukowski}, {\it {Violations of local
  realism with quantum systems described by N-dimensional Hilbert spaces up to
  N=16}},  {\em PhysRevA} {\bf 64} (Aug., 2001) 024101,
  [\href{http://arxiv.org/abs/quant-ph/0101084}{{\tt quant-ph/0101084}}].

\bibitem{2211.05795}
K.~{H{\"a}ring}, A.~{Hebbar}, D.~{Karateev}, M.~{Meineri}, and J.~{Penedones},
  {\it {Bounds on photon scattering}},  {\em arXiv e-prints} (Nov., 2022)
  arXiv:2211.05795, [\href{http://arxiv.org/abs/2211.05795}{{\tt
  arXiv:2211.05795}}].

\bibitem{1810.06994}
J.~{Quevillon}, C.~{Smith}, and S.~{Touati}, {\it {The effective action for
  gauge bosons}},  {\em arXiv e-prints} (Oct., 2018) arXiv:1810.06994,
  [\href{http://arxiv.org/abs/1810.06994}{{\tt arXiv:1810.06994}}].

\bibitem{Heisenberg}
W.~Heisenberg and H.~Euler, {\it {Consequences of Dirac's theory of
  positrons}},  {\em Z. Phys.} {\bf 98} (1936), no.~11-12 714--732,
  [\href{http://arxiv.org/abs/physics/0605038}{{\tt physics/0605038}}].

\bibitem{1202.1557}
G.~V. {Dunne}, {\it {The Heisenberg-Euler Effective Action: 75 Years on}},
  {\em International Journal of Modern Physics A} {\bf 27} (June, 2012)
  1260004, [\href{http://arxiv.org/abs/1202.1557}{{\tt arXiv:1202.1557}}].

\bibitem{2012.05798}
L.~{Alberte}, C.~{de Rham}, S.~{Jaitly}, and A.~J. {Tolley}, {\it {QED
  positivity bounds}},  {\em PhysRevD} {\bf 103} (June, 2021) 125020,
  [\href{http://arxiv.org/abs/2012.05798}{{\tt arXiv:2012.05798}}].

\bibitem{WGC}
N.~{Arkani-Hamed}, L.~{Motl}, A.~{Nicolis}, and C.~{Vafa}, {\it {The string
  landscape, black holes and gravity as the weakest force}},  {\em Journal of
  High Energy Physics} {\bf 2007} (June, 2007) 060,
  [\href{http://arxiv.org/abs/hep-th/0601001}{{\tt hep-th/0601001}}].

\bibitem{2103.12728}
Z.~{Bern}, D.~{Kosmopoulos}, and A.~{Zhiboedov}, {\it {Gravitational effective
  field theory islands, low-spin dominance, and the four-graviton amplitude}},
  {\em Journal of Physics A Mathematical General} {\bf 54} (Aug., 2021) 344002,
  [\href{http://arxiv.org/abs/2103.12728}{{\tt arXiv:2103.12728}}].

\bibitem{schmidt}
P.~Kaye, R.~Laflamme, and M.~Mosca, {\em An Introduction to Quantum Computing}.
\newblock 11, 2006.

\end{thebibliography}\endgroup

\end{document}